\documentclass[twocolumn]{aastex63}
\usepackage{latexsym}
\usepackage{amsmath}
\usepackage{natbib}
\usepackage{graphicx}
\usepackage{mathrsfs}
\usepackage{color}

\newcommand{\hi}{H\textsc{i}}

\def\kms{km~s$^{-1}$}
\def\cmcu{cm$^{-3}$}

\shorttitle{Tiny-scale Structure Discovered toward PSR~B1557$-$50}
\shortauthors{Liu et al.}

\graphicspath{{./}{figures/}}
\begin{document}

\title{Tiny-scale Structure Discovered toward PSR~B1557$-$50}

\correspondingauthor{Mengting Liu, Di Li}
\email{liumengting@nao.cas.cn}
\email{dili@nao.cas.cn}

\author[0000-0001-7790-5498]{Mengting Liu}
\affil{National Astronomical Observatories, Chinese Academy of Sciences, Beijing 100101, People's Republic of China}
\affil{University of Chinese Academy of Sciences, Beijing 100049, People's Republic of China}

\author[0000-0001-9659-0292]{Marko Kr\v{c}o}
\affil{National Astronomical Observatories, Chinese Academy of Sciences, Beijing 100101, People's Republic of China}

\author[0000-0003-3010-7661]{Di Li}
\affil{National Astronomical Observatories, Chinese Academy of Sciences, Beijing 100101, People's Republic of China}
\affil{University of Chinese Academy of Sciences, Beijing 100049, People's Republic of China}
\affil{NAOC-UKZN Computational Astrophysics Centre, University of KwaZulu-Natal, Durban 4000, South Africa }

\author[0000-0003-1502-100X]{George Hobbs}
\affil{CSIRO Astronomy $\&$ Space Science, Australia Telescope National Facility, P.O. Box 76, Epping, NSW 1710, Australia}

\author[0000-0003-0235-3347]{J.\ R.\ Dawson}
\affil{CSIRO Astronomy $\&$ Space Science, Australia Telescope National Facility, P.O. Box 76, Epping, NSW 1710, Australia}
\affiliation{Department of Physics and Astronomy and MQ Research Centre in Astronomy, Astrophysics and Astrophotonics, Macquarie University, Sydney, NSW 2109, Australia}

\author[0000-0002-7456-8067]{Carl Heiles}
\affil{Department of Astronomy, University of California, Berkeley, CA 94720-3411, USA}

\author[0000-0002-0996-3001]{Andrew Jameson}
\affil{Centre for Astrophysics and Supercomputing, Swinburne University of Technology, P.O. Box 218, Hawthorn, Victoria 3122, Australia}
\affil{Australia Research Council Centre for Excellence for Gravitational Wave Discovery (OzGrav)}

\author[0000-0002-3418-7817]{Sne\v{z}ana Stanimirovi\'c}
\affil{Department of Astronomy, University of Wisconsin–Madison, Madison, WI 53706, USA}

\author[0000-0002-7122-4963]{Simon Johnston}
\affiliation{CSIRO Astronomy $\&$ Space Science, Australia Telescope National Facility, P.O. Box 76, Epping, NSW 1710, Australia}

\author[0000-0002-6300-7459]{John M. Dickey}
\affil{School of Maths and Physics, University of Tasmania, Hobart, TAS 7001, Australia}


\begin{abstract}

Optical depth variations in the Galactic neutral interstellar medium (ISM) with spatial scales from hundreds to thousands of astronomical units have been observed through \hi\ absorption against pulsars and continuum sources, while extremely small structures with spatial scales of tens of astronomical units remain largely unexplored. 
The nature and formation of such tiny-scale atomic structures (TSAS) need to be better understood. 
We report a tentative detection of TSAS with a signal-to-noise ratio of 3.2 toward PSR~B1557$-$50 in the second epoch of two Parkes sessions just 0.36\,yr apart, which are the closest-spaced spectral observations toward this pulsar.
One absorption component showing marginal variations has been identified. Based on the pulsar's proper motion of 14\,mas $\rm yr^{-1}$ and the component's kinematic distance of 3.3\,kpc, the corresponding characteristic spatial scale is 17\,au, which is among the smallest sizes of known TSAS.
Assuming a similar line-of-sight (LOS) depth, the tentative TSAS cloud detected here is
overdense and overpressured relative to the cold neutral medium (CNM), and
can radiatively cool fast enough to be in thermal equilibrium with the ambient environment. We find that turbulence is not sufficient to confine the overpressured TSAS.
We explore the LOS elongation that would be required for the tentative TSAS to be at the canonical CNM pressure, and find that it is $\sim5000$---much larger than filaments observed in the ISM.
We see some evidence of
line width and temperature variations in the CNM components observed at the two epochs, as predicted by models of TSAS-like cloud formation colliding warm neutral medium flows.

\end{abstract}


\section{Introduction} \label{sec:intro}

Tiny-scale atomic structures (TSAS) with spatial scales spanning from a few to hundreds of astronomical units have been studied for over four decades in the interstellar medium (ISM).
Compared to the traditional definitions of the atomic medium in thermal equilibrium states---the cold neutral medium (CNM) and the warm neutral medium (WNM)---TSAS are overdense and overpressured, with densities and pressures several orders of magnitude higher than the theoretical values based on the heating and cooling balance \citep{2018ARA&A..56..489S}.
 TSAS has been probed mainly through three methods:
(1) spatial and temporal $\rm \hi$ absorption variability from high-resolution maps against extragalactic compact and resolved radio continuum sources (e.g. \citealp{1989ApJ...347..302D,2000ApJ...543..227D,2009AJ....137.4526L,2012ApJ...749..144R,2020ApJ...893..152R});
(2) temporal variability of $\rm \hi$ absorption against pulsars  (e.g. \citealp{1994ApJ...436..144F, 2003MNRAS.341..941J,2005ApJ...631..376M,2010ApJ...720..415S});
(3) spatial and temporal absorption variability of optical and ultraviolet lines (e.g. {Na\,{\sc i}}, {Ca\,{\sc i}}) toward stars in binary, multiple stellar systems, and globular clusters over a period of years (e.g. \citealp{1990ApJ...364L...5M,1996ApJ...464L.179M,2003ApJ...591L.123L}). 

Several mechanisms have been proposed to form TSAS, although there is no consensus yet. These mechanisms include TSAS being the tail-end of the turbulent spectrum\citep{2000ApJ...543..227D}, discrete elongated disk or cylinder clouds passing across the line of sight \citep{1997ApJ...481..193H}, isolated events, such as fragmentation of the Local Bubble wall through hydrodynamic instabilities \citep{2010ApJ...720..415S}, and structures related to planet-building gas materials \citep{2017ApJ...836..135R,2018ARA&A..56..489S}.
If TSAS represents the tail-end of turbulent dissipation processes, then a power-law relation between the spatial scales of TSAS and the optical depth variation is expected \citep{2000MNRAS.317..199D}. 
Alternatively, \citet{2002ApJ...564L..97K} and \citet{2007A&A...465..431H} modeled TSAS creation by shocks and colliding WNM flows in shocked regions.

Pulsars are particularly useful background sources because:
(1) on- and off-source measurements can be obtained without position switching; (2) the relative high-velocity motion of the Earth and a pulsar over time allows us to sample a slightly different line-of-sight(LOS) through the ISM, therefore probing small structures in the ISM; (3) \hi\ absorption spectra of low-latitude pulsars can be used to estimate the pulsar kinematic distance \citep{2008ApJ...674..286W};

TSAS may contain ionized gas. For example, assuming shielding is not too high, a high fraction of the carbon in TSAS clouds may be ionized by the interstellar radiation field (\citealt{1997ApJ...481..193H, 2018ARA&A..56..489S}). 
Ionization of hydrogen may also be instigated by local stellar objects, to produce fully ionized tiny-scale structures.
\citet{2001AJ....122.1500M} detected an extremely narrow ionized filament, which can be attributed to the effect of photoionization from a star or compact object. 
Tiny-scale inhomogeneities in the ionized ISM can also be discerned through scintillation and/or extreme scattering events \citep{2018ARA&A..56..489S}. For example, \citet{2016MNRAS.458.2509B}, \citet{2015ApJ...808..113C}, and \citet{2005ApJ...619L.171H} detected tiny-scale ionized structure with spatial scales of a few to hundreds of astronomical units through extreme scattering events and substructure in scintillation arcs toward pulsars.

PSR~B1557$-$50 (J1600$-$5044) was the first pulsar against which \hi\ absorption variations were detected in the southern sky \citep{1992MNRAS.258P..19D}. 
In observations spaced around five years apart, these authors found an overpressured TSAS with a spatial scale of 1000 au, at a velocity of $-$40 \kms\ with $\Delta\tau$ of 1.1. 
\citet{2003MNRAS.341..941J} confirmed similar \hi\ absorption variation at $-$40 \kms\ as well as another two components at $-$110 and $-$80 \kms\ toward this pulsar using Parkes, in two observational epochs spaced 14 yr apart. They concluded that the component at $-$110 \kms\ is related to a TSAS of 1000 au, with a density of 2.6$\times$10$^4$ \cmcu, and $\Delta\tau$ of 0.15. 
The sight line toward PSR~B1557$-$50 traverses substantial distances, and has previously been monitored at such long cadences that multiple TSAS components could have passed across the source between observations. 
We focus here on monitoring PSR~B1557$-$50 over a much shorter time baseline: $\sim$0.36 yr.

This Letter is organized as follows:  Section \ref{sec:obs} describes the observing and data processing strategies. In Section \ref{sec:tsas_prop_lpv}, we decompose the optical depth profile and study the properties of the CNM.
Meanwhile, we analyze the variations of \hi\ optical depth profiles and the properties of the changes in the observed \hi\ gas at different epochs. 
We discuss our results and their implications in Section \ref{sec:disc}. We conclude in 
Section \ref{sec:conc}.

\section{OBSERVATIONS AND DATA PROCESSING} \label{sec:obs}

We used the Parkes ultra-wide-bandwidth, low-frequency receiver (`UWL') and the MEDUSA signal processing system \citep{2020PASA...37...12H}, to observe the 21cm \hi\ line towards PSR~B1557$-$50 at two epochs, referred to below as E1(2019.42) and E2(2019.78). The UTC dates at the beginning of the observations are 2019 May 26 and 2019 September 11, respectively. 
The integration time for each epoch was
379 and 382 minutes, respectively. 

We recorded voltage data streams covering frequencies from 1344 to 1472 MHz.
We injected a calibration signal (CAL) every 1 hr for a duration of 3 minutes in order to calibrate the antenna temperature. 
We used the {\tt\string DSPSR} package \citep{2011PASA...28....1V} to process the baseband data to get spectra with a velocity resolution of $\sim$0.1 \kms.
The data were folded synchronously with the pulsar period for an integration time of 10 s with 64 phase bins. 

The pulse profile for each polarization is generated by collapsing (after de-dispersion) the data along the frequency dimension, which gives the location of pulsar-on and pulsar-off phase bins.
The pulsar-on and pulsar-off spectra were obtained based on the strategies described by \citet{2008ApJ...674..286W}. 
The absorption spectrum is the difference of the pulsar-on and pulsar-off spectra on an antenna temperature scale, $I(v) = I^{\rm on} (v)-I^{\rm off} (v)$.
The final normalized absorption spectrum ($I(v)/I_0$) was obtained by dividing the final absorption spectrum by the mean value excluding the absorption channels ($I_0$).
According to 
radiative transfer, the pulsar-on and pulsar-off spectra can be written as 
\begin{equation}
\begin{split}
T^{\rm on} (v)&= (T_{\rm bg}+T_{\rm psr})e^{-\tau_v}+T_{\rm s}(1-e^{-\tau_v}),\\
T^{\rm off} (v)&= T_{\rm bg}e^{-\tau_v}+T_{\rm s}(1-e^{-\tau_v}),
\end{split}
\label{e:on}
\end{equation}
\noindent where $T_{\rm bg}$ is the background brightness temperature, $\tau_v$ is the \hi\  optical depth, $T_{\rm s}$ is the \hi\ spin temperature, and $T_{\rm psr}$ is the pulsar continuum brightness temperature.
Therefore,
\begin{equation}
e^{-\tau_v}= (T^{\rm on} (v)-T^{\rm off} (v))/T_{\rm psr}=I(v)/I_0.
\label{e:tau}
\end{equation}

The final calibrated \textit{antenna} temperature of the $\rm \hi$ emission spectrum, $T_{\rm HI}(v)$, is required to compute the frequency-dependent noise spectrum (see \S\ref{sec:line_prof}). 
$T_{\rm HI}(v)$ was obtained by removing the baseline of $I^{\rm off}(v)$ using a third-order polynomial.
To obtain the 
brightness temperature-calibrated $T^{\rm off} (v)$, which is required to derive $T_{\rm s}$ (see Appendix \ref{sec:gs}), we scaled $T_{\rm HI}(v)$ to match the Parkes Galactic All-Sky Survey (GASS) intensity scale \citep{2009ApJS..181..398M}.

The absorption spectra are affected by on-source ripples, especially for E2. 
We identified the Fourier components of the ripples. Among these there are two baseline ripples with lag times of 0.175 and 0.35 $\mu$s (the second harmonic ripple) corresponding to the periods of standing waves formed by the reflections between the vertex and the focal plane. We did not attempt to diagnose formation mechanisms for any other ripples.
All ripples were reduced in $I(v)$ by fitting for and removing sinusoidal functions with the same Fourier components.

\begin{figure*}
\begin{center}
\includegraphics[width=1.0\textwidth]{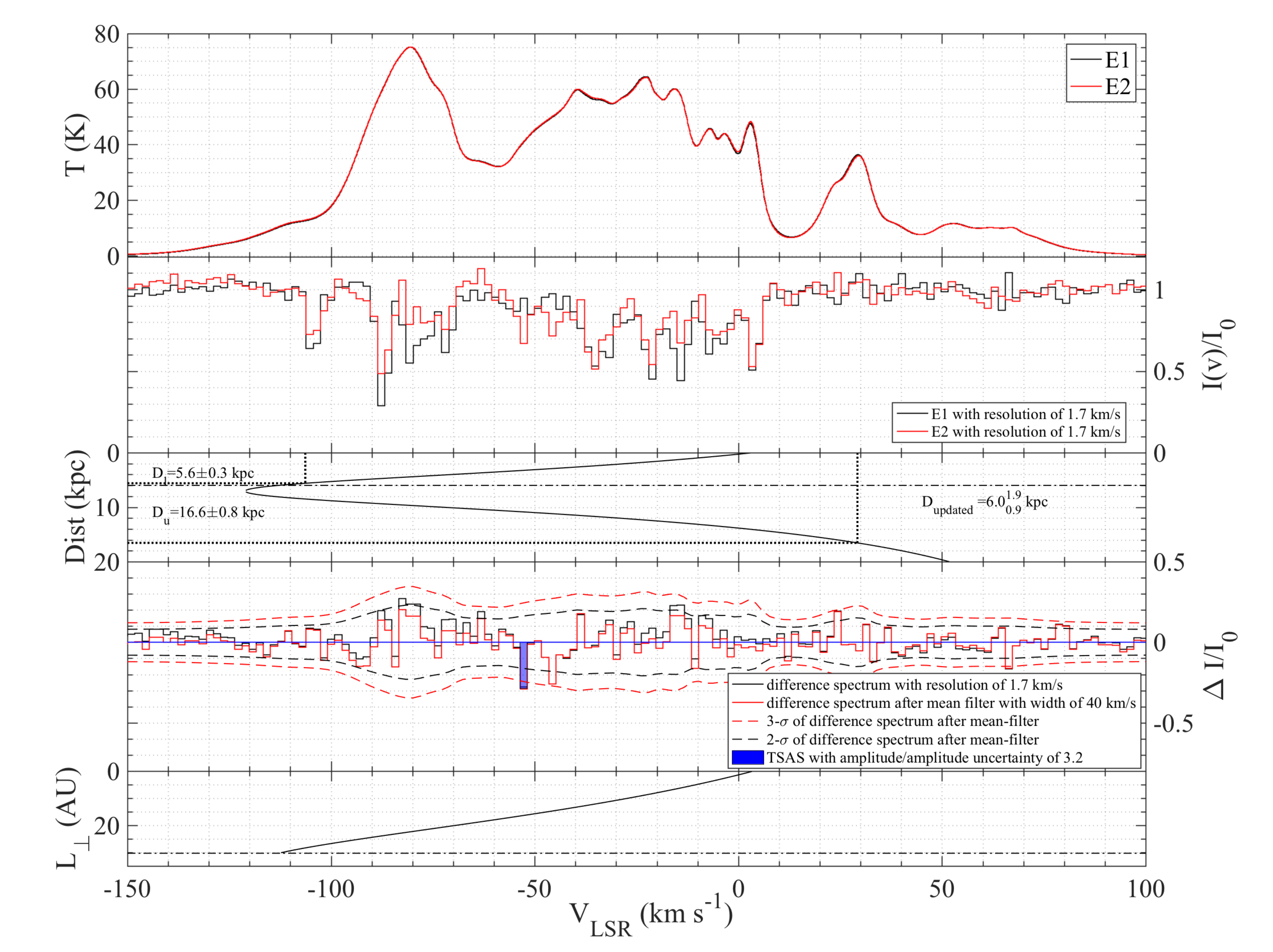}
\caption{ Spectra for PSR~B1557$-$50. First panel: \hi\ emission scaled to match GASS in the direction of the pulsar with a velocity resolution of 0.1 \kms. 
Second panel: 
Velocity-binned HI absorption spectra taken at E1 (black) and E2 (red) with a channel width of $\sim$1.7 \kms. $I(v)/I_0$ is as defined in Equation \ref{e:tau}.
Third panel: distance versus radial velocity curve, calculated from \citet{2019ApJ...870L..10M} using the best-fit linear model with $R_0$=8.09\,kpc, $\Theta_0$= 233.6 \kms.
Fourth panel: 
velocity-binned difference \hi\ absorption profiles ($\Delta I(v) /I_{0} = [I(v)/I_0]_{\rm E2} - [I(v)/I_0]_{\rm E1}$) with a channel width of 1.7 \kms, before (black) and after (red) mean-filtering. Red and black dashed lines represent $\pm$ 2 and 3$\sigma$ noise envelopes where the contribution from \hi\ emission has been taken into account. The blue channel represents the TSAS we have identified with an S/N of 3.2.
The blue horizontal line represents 0 difference.
Fifth panel: spatial scale as a function of velocity.
The dashed-dotted line represents the maximum spatial scale corresponding to the transverse distance traveled by the pulsar between the two epochs.}
\label{f:abs}
\end{center}
\end{figure*}

The \hi\ emission and absorption spectra from E1 and E2 are shown in the first and second panels of Figure \ref{f:abs}, respectively. 
There are clear differences in the absorption features between the two epochs, which we discuss further below.

Below the spectra, we display the distance versus radial velocity curve, which was derived for the Galactic coordinates of PSR~B1557$-$50 ($l=$330.690$^{\circ}$, $b=$1.631$^{\circ}$) using a linear rotation curve \citep{2019ApJ...870L..10M}.
A detailed description of the derivation is given in Appendix \ref{sec:d_rv}.
The lower distance limit $D_{l}$=$5.6 \pm 0.3$\,kpc for PSR~B1557$-$50 is set by the velocity center of the most distant absorption feature.
The upper distance limit $D_{u}=16.6\pm 0.8$\,kpc
is given by the distance of the nearest significant emission component \citep[brightness temperature above 35\,K;][]{1979A&A....77..204W} not seen in absorption. 
We adopted the method described in \citet{2012ApJ...755...39V} to translate these lower and upper limits into actual distance using a likelihood analysis. 
The updated pulsar distance $D_{\rm updated}$ is $6.0^{1.8}_{0.6}$\,kpc. 
This is lower than the distance from the Australia Telescope National Facility (ATNF) pulsar catalog $D_{\rm psrcat} = 6.9^{1.9}_{0.9}$\,kpc, which was estimated by \citet{2012ApJ...755...39V} based on \hi\ distance limits from \citet{2001MNRAS.322..715J} with a flat rotation curve \citep{1989ApJ...342..272F}, while higher than the distance determined from recent electron density models, which give $D_{\rm DM}$=5.1\,kpc \citep{2017ApJ...835...29Y}.

\section{Spectral Analysis, Line Profile Variations, and TSAS Properties} \label{sec:tsas_prop_lpv}

\subsection{Spectral Analysis} \label{sec:spin}

We largely followed the method of \citet{2018ApJS..238...14M} to estimate the physical properties of each \hi\ component, assuming that CNM can be seen in both the optical depth spectrum and the \hi\ emission profile, while the WNM can only be seen in the emission profile. All CNM and WNM components were obtained through Gaussian decomposition \citep{2003ApJS..145..329H}.
A detailed description of the spectral decomposition and spin temperature calculation is given in Appendix \ref{sec:gs}.

The derived spin temperatures for each CNM component in both epochs are listed in Table \ref{tab:params}.
Most of the CNM clouds from the two observations have spin temperatures in the range 10--260\,K, which is consistent with previous observed CNM clouds \citep{2003ApJ...586.1067H}.

\begin{deluxetable*}{cccc|cc}
\setlength{\tabcolsep}{0.05in} 
\tablecolumns{11} 
\tabletypesize{\scriptsize}
\tablewidth{0pt}
\tablecaption{Fitted Parameters of CNM Components from Gaussian Decomposition\label{tab:params}}
\tablehead{
\colhead{Epoch}  &  \colhead{$\tau_0$} &  \colhead{$v_0$}         &  \colhead{$\Delta v_0$}  &    \colhead{$T_s$}       & \colhead{$N({\rm HI})_{\rm abs}$}     \\ [1mm]
\colhead{(name)}   &          &  \colhead{($\rm km\,s^{-1}$)}  & \colhead{($\rm km\,s^{-1}$)}   &    \colhead{($\rm K$)}  & \colhead{($10^{20}\rm\,cm^{-2}$)} }
\startdata
E1 (2019.42) &	0.9	$\pm$	0.1	&	$-$106.18	$\pm$	0.09	&	1.6	$\pm$	0.2	&	11.5	$\pm$	3.1	&	0.3	$\pm$	0.1	\\
&	3.3	$\pm$	0.1	&	$-$88.25	$\pm$	0.02	&	1.49	$\pm$	0.06	&	45.7	$\pm$	3.2	&	4.4	$\pm$	0.4	\\
&	1.0	$\pm$	0.1	&	$-$81.45	$\pm$	0.08	&	0.9	$\pm$	0.2	&	8.1	$\pm$	4.9	&	0.1	$\pm$	0.1	\\
&	0.35	$\pm$	0.06	&	$-$79.6	$\pm$	0.8	&	8.4	$\pm$	2.2	&	203.7	$\pm$	24.5	&	12	$\pm$	4.1	\\
&	0.5	$\pm$	0.1	&	$-$72.8	$\pm$	0.3	&	2.6	$\pm$	0.7	&	97.0	$\pm$	7.3	&	2.2	$\pm$	0.8	\\
&	0.67	$\pm$	0.05	&	$-$35.7	$\pm$	0.2	&	5.9	$\pm$	0.6	&	105.6	$\pm$	4.7	&	8.3	$\pm$	1.1	\\
&	0.79	$\pm$	0.07	&	$-$22.7	$\pm$	0.2	&	3.5	$\pm$	0.4	&	103.7	$\pm$	3.8	&	5.7	$\pm$	0.8	\\
&	0.83	$\pm$	0.07	&	$-$15.9	$\pm$	0.2	&	3.3	$\pm$	0.4	&	96.2	$\pm$	4.0	&	5.3	$\pm$	0.8	\\
&	0.5	$\pm$	0.05	&	$-$6.6	$\pm$	0.3	&	6.2	$\pm$	0.8	&	98.7	$\pm$	5.7	&	6.2	$\pm$	1.1	\\
&	0.79	$\pm$	0.08	&	2.0	$\pm$	0.1	&	2.8	$\pm$	0.3	&	137.7	$\pm$	9.1	&	6.0	$\pm$	1.0	\\	
\hline
E2 (2019.78)
&	0.5	$\pm$	0.07	&	$-$105.3	$\pm$	0.2	&	2.4	$\pm$	0.4	&	10.3	$\pm$	2.5	&	0.25	$\pm$	0.08	\\
&	1.1	$\pm$	0.1	&	$-$88.61	$\pm$	0.06	&	1.4	$\pm$	0.1	&	16.7	$\pm$	11.5	&	0.6	$\pm$	0.4	\\
&	0.25	$\pm$	0.04	&	$-$78.1	$\pm$	0.7	&	7.4	$\pm$	1.8	&	201.8	$\pm$	37.7	&	7.3	$\pm$	2.6	\\
&	0.4	$\pm$	0.1	&	$-$72.5	$\pm$	0.1	&	1.2	$\pm$	0.4	&	17.6	$\pm$	7.7	&	0.2	$\pm$	0.1	\\
&	0.37	$\pm$	0.06	&	$-$53.99	$\pm$	0.3	&	3.7	$\pm$	0.7	&	75.5	$\pm$	13.6	&	2.0	$\pm$	0.7	\\
&	0.32	$\pm$	0.05	&	$-$46.6	$\pm$	0.4	&	4.7	$\pm$	1.1	&	151.6	$\pm$	10.5	&	4.6	$\pm$	1.3	\\
&	0.5	$\pm$	0.1	& $-$37.3	$\pm$	0.2	&	2.1	$\pm$	0.6	&	14	$\pm$	7.9	&	0.3	$\pm$	0.2	\\
&	0.38	$\pm$	0.06	&	$-$34.8	$\pm$	0.7	&	8.3	$\pm$	1.3	&	157.4	$\pm$	10.2	&	9.8	$\pm$	2.3	\\
&	0.58	$\pm$	0.06	&	$-$22.5	$\pm$	0.2	&	4.3	$\pm$	0.5	&	86.9	$\pm$	14.5	&	4.3	$\pm$	1.0	\\
&	0.43	$\pm$	0.06	&	$-$15.0	$\pm$	0.2	&	3.2	$\pm$	0.6	&	131.7	$\pm$	7.0	&	3.6	$\pm$	0.9	\\
&	0.35	$\pm$	0.04	&	$-$5.8	$\pm$	0.5	&	7.1	$\pm$	1.2	&	99.5	$\pm$	19.8	&	5.0	$\pm$	1.4	\\
&	0.68	$\pm$	0.07	&	2.3	$\pm$	0.2	&	3.0	$\pm$	0.4	&	102.1	$\pm$	20.7	&	4.1	$\pm$	1.1	\\
\enddata
\tablecomments{Column (1): observing epoch. Columns (2-4): gaussian parameters fit to the opacity profile (Equation~\ref{e:tau_abs}). Column (5): \hi\ spin temperature. Column (6): \hi\ column density of absorption component calculated by
$N({\rm H\textsc{i}})_{\rm abs} =  C_0 \int \tau \, T_s \, dv  = 1.064 \cdot C_0 \cdot \tau_0 \cdot \Delta v_0 \cdot T_s$, where $C_0 = 1.823\times 10^{18} \rm\,cm^{-2} / (km\,s^{-1}\,K)$ \citep{2018ApJS..238...14M}.
}
\end{deluxetable*}

\subsection{Line Profile Variations and TSAS properties}
\label{sec:line_prof}

In order to evaluate H\textsc{i} absorption profile variations between the two epochs, we first define a difference H\textsc{i} absorption profile $\Delta I(v) /I_{0} = [I(v)/I_0]_{\rm E2} - [I(v)/I_0]_{\rm E1}$.
Here it is crucial to account accurately for the noise profile, where the \hi\ emission tends to dominate over system temperature in this band \citep{2008ApJ...674..286W}. We thus added the \hi\ antenna temperature, in the velocity range where it can be measured, to the system temperature.
The resultant noise profile is given by $\sigma_{\Delta I/I_{0}}(v) = \sigma_{\Delta I /I_{0},\rm off-line} \times [T_{\rm HI}(v) + T_{\rm sys,off-line}]/T_{\rm sys,off-line}$, where $T_{\rm sys,off-line}$ is the system temperature for off-line channels, $T_{\rm HI}(v)$ is the antenna temperature of \hi\ at velocity $v$, and $\sigma_{\Delta I /I_{0},\rm off-line}$ is the absorption spectrum standard deviation of off-line channels \citep{2008ApJ...674..286W}. 
In order to mitigate the effect of residual baseline ripples on $\Delta I(v)/I_{0}$, we constructed a boxcar mean-filter with a width of 40 \kms\ , based on the width of the residual ripples, and subtracted it from the difference spectrum.
We propagated the uncertainties associated with the mean-filter to determine the final noise envelope of the difference spectrum after mean-filtering.

We adopted a two-stage approach to Gaussian fitting of the mean-filtered difference spectrum. First, we used a Markov Chain Monte
Carlo (MCMC) method to obtain an initial list of TSAS candidates. 
The MCMC sampling was performed using the {\tt\string emcee} package \citep{2013PASP..125..306F}, with the range of Gaussian parameters for the prior obtained from the properties of the absorption components at the two epochs.
We then treated all components with local minima as constrained from MCMC as TSAS candidates.
We next performed a traditional $\chi^{2}$ fitting on those candidates using the {\tt\string LMFIT} package \citep{matt_newville_2019_3381550}, 
with initial guesses for the Gaussian parameters drawn from the MCMC output.
We define the signal-to-noise ratio (S/N) of the component as the ratio of the fitted amplitude to the 1$\sigma$ amplitude uncertainty output by LMFIT (where the appropriate noise spectrum has been used as the uncertainty input). This 1$\sigma$ amplitude uncertainty is estimated from the square root of the diagonal elements of the covariance matrix.
By this definition, there is one tentative TSAS showing a marginal detection with an S/N of 3.2 at a velocity of $\sim-$54 \kms, 
shown in the fourth panel of Figure \ref{f:abs}. Note that while the spectrum is
binned in velocity to a channel width of 1.7 \kms\ to illustrate the TSAS clearly, the fitting was performed on the unbinned data.

Compared to previous TSAS detections toward this pulsar (\citealt{1992MNRAS.258P..19D,2003MNRAS.341..941J}), we have a much higher velocity resolution of $\sim$0.1 \kms, reduced artifacts on the difference spectrum, and an estimated distance for each channel, enabling us to obtain TSAS detection results with definitive spatial scales. The spatial scale $L_{\perp}(v)$ was calculated using $L_{\perp}(v)=\pi\rm PM_{TOT}/\left(3600*180 \right)\times Dist \times\Delta t$, where $\rm PM_{TOT}$ is pulsar total proper motion of 14\,mas\,yr$^{-1}$ from the ATNF pulsar catalog \citep{2005AJ....129.1993M}, and $\Delta t$ is the time baseline.
 The characteristic scale of the tentative TSAS that we probe is 17\,au, which is much smaller than the TSAS previously detected toward PSR~B1557$-$50 ($\sim$1000\,au at the velocity of $-$110 \kms\ assuming the pulsar velocity of 400 \kms\ and the component distance of 6.4 kpc) with a time baseline of 20 yr.
This is due to the short time interval between the two observations.

The observed properties of the single tentatively detected TSAS component are summarized in Table \ref{tab:comp_pro}. 
We assume that the TSAS has the same spin temperature as its host CNM cloud ($75.5\pm13.6$\,K). 
The derived value of the \hi\ column density of the  tentative TSAS is ($7\pm3$)$\times 10^{19} \rm\,cm^{-2}$, consistent with the column densities of previous TSAS studies, which range from 10$^{19}$ to 10$^{21} \rm\,cm^{-2}$ \citep{2018ARA&A..56..489S}.
In total, the tentative TSAS feature contributes $\sim 34\%$ to the total \hi\ column density of its host CNM cloud.

\section{Discussion}
\label{sec:disc}

If the tentative TSAS is a spherical cloud with diameter of $L_{\perp}$, 
the derived \hi\ volume density and thermal pressure are larger than $\sim 10^4$\,cm$^{-3}$ and $\sim 10^6$\,K\,cm$^{-3}$, respectively.
These values are consistent with previous TSAS studies (assuming similar geometry), with TSAS showing densities and thermal pressures two to three orders of magnitude larger than typical ISM values \citep{1997ApJ...481..193H,2010ApJ...720..415S,2018ARA&A..56..489S}.

\begin{deluxetable*}{cccccccccccccccccc}
\setlength{\tabcolsep}{0.05in} 
\tabletypesize{\scriptsize}
\tablewidth{0pt}
\tablecaption{Tentative TSAS properties \label{tab:comp_pro} }
\tablehead{
\colhead{Center Velocity}      &\colhead{ $L_{\perp}$ }& \colhead{Dist}& \colhead{$\Delta v_0$} &  \colhead{$\Delta\tau$} &\colhead{$T_s$} &
\colhead{$N(\rm HI)_{\rm TSAS}$}    & \colhead{$n(\rm HI)_{\rm TSAS}$ } & $P/k$ &\colhead{$N_{\rm min,c}$}&\colhead{$\sigma_{\rm tur}$/$\sigma_{\rm tur}^{T_s=10\,K}$}& \colhead{$\sigma_{\rm th}$/$\sigma_{\rm th}^{T_s=10\,K}$}\\
\colhead{(\kms)}   &\colhead{(au)}& \colhead{(kpc)}   &   \colhead{(\kms)} &\colhead{}&
\colhead{($\rm\,K$) } &     \colhead{($10^{20}\rm\,cm^{-2}$) }&  \colhead{($10^{4}\rm\,cm^{-3}$) } &  \colhead{($10^{6}\rm\,cm^{-3}\,K$) } &     \colhead{($10^{20}\rm\,cm^{-2}$) }&     \colhead{(\kms) }&
\colhead{(\kms) }
}
\startdata
$-53.5\pm 0.2$  & 17 & 3.3$\pm$0.3& 1.3 $\pm$0.5  
&  0.35$\pm$0.11 & 75.5$\pm$13.6 & 0.7$\pm$0.3  & 26.6 $\pm$13.7 & 20.7$\pm$12.5  & 0.03 &0.2$\pm$0.1/0.5 & 0.8$\pm$0.1/0.3\\  
\enddata
\tablecomments{ Column (1): TSAS central velocity. Column (2): TSAS spatial scale. Column (3): TSAS distance. Column (4): FWHM of TSAS. Column (5) maximum absolute value of optical depth variations.  Column (6): spin temperature. 
Column (7): TSAS \hi\ column density. It is calculated by
$N{(\rm HI)}_{\rm TSAS} = 1.064 \cdot C_0 \cdot |\Delta\tau| \cdot \Delta v_0 \cdot T_s$. 
Column (8): TSAS \hi\ volume density. Column (9): thermal pressure. Column (10): the minimum column density calculated when the dynamical time is larger than the cooling time. Column (11): the one-dimensional turbulent velocity/ the one-dimensional turbulent velocity when the TSAS spin temperature is 10\,K. Column (12): the one-dimensional thermal velocity/ the one-dimensional thermal velocity when the TSAS spin temperature is 10\,K.} 
\end{deluxetable*}

When the dynamical time $t_{dyn}$ is larger than the cooling time $t_{cool}$, clouds may reach a thermal equilibrium, even if they are overpressured. 
The column density $N_{\rm min,c}$ for a cloud at the balance of cooling and expanding can be calculated.
As estimated by \citet{2018ARA&A..56..489S},
\begin{equation}
\begin{split}
t_{dyn}&>t_{cool},\\
\dfrac{R}{\sqrt{kT/m}} &> \dfrac{3kT}{2n \Lambda},\\
nR &> N_{\rm min,c}= 1.2 \times 10^{15} T^{3/2} e^{92/T},
\end{split}
\label{e:N_min}
\end{equation}
where $R$ is the size of the cloud, $T$ is the kinetic temperature, and $\Lambda$ is the radiative loss function assuming the C\textsc{ii} fine structure line is the main source of cooling, with a carbon depletion
factor of 0.35.
Clouds with a column density larger than $N_{\rm min,c}$ would cool radiatively faster than they expand and reach thermal equilibrium. We use the \hi\ spin temperature as an approximation of the kinetic temperature \citep{2003ApJ...591L.123L} to estimate $N_{\rm min,c}$.
We found that our tentative TSAS component has $N{(\rm HI)}_{\rm TSAS}$ significantly larger than $N_{\rm min,c}$, suggesting that it may be overpressured and in thermal equilibrium with the ambient ISM.

To alleviate the problem of overpressurization, \citet{1997ApJ...481..193H} 
proposed a model in which TSAS represents discrete features in the shape of cylinders and disks that are
homogeneously and isotropically distributed within CNM clouds.
The volume density of TSAS depends both on the spatial scale across the LOS, $L_{\perp}$, and the path length along the LOS $L_{\parallel}$. 
Following \citet{1997ApJ...481..193H}, we define the geometric elongation factor $\mathscr{G}$ = $L_{\parallel}/$$L_{\perp}$ and use the standard CNM thermal pressure of 4000 $\rm\,cm^{-3}\,K$.
Based on our measured spin temperature and column density, the elongation factor $\mathscr{G}$ required for our tentative TSAS to match the CNM pressure is $\sim 5000$. 
Such extremely filamentary structure
is out of the range predicted by \citet{1997ApJ...481..193H}, who found $\mathscr{G}$ larger than 1 and less than 10.
This is a result of the large variation of optical depth and the small spatial scale of the tentative TSAS that we detected.

\citet{2007A&A...465..431H} 
found that CNM clouds can be generated from thermally unstable regions 
in colliding WNM flows based on simulations with resolutions of 400 and 4000 au (larger than the typical spatial scales of observed TSAS). 
The supersonic collisions of CNM clouds form transient shocked regions with large temperature variations at the boundaries, showing similar properties to TSAS. Such temperature variations could induce line width and optical depth variations in the observed \hi\ absorption spectrum.
From Figure \ref{f:abs} and Table \ref{tab:params} it can be seen that, in addition to apparent optical depth variations, the spin temperatures and line widths of the
CNM components 
detected in this work do exhibit 
differences between the two epochs.
While much of this variation lies below strict $3\sigma$ limits, it may still be indicative of the variations predicted by colliding flows models, possibly suggesting that our tentative TSAS cloud could have formed from such colliding WNM flows. However, better evidence of variability in the data, together with detailed descriptions of the properties of TSAS from higher-resolution simulations (\textless 100 au) are needed to make firm quantitative comparisons.

Finally, compressible turbulence can generate density fluctuations and may even be able to confine an overpressured structure.
\citet{2018ARA&A..56..489S} estimate that in order to 
confine TSAS-like structures that are overpressured by a factor of $\sim$100, the turbulent velocity must be at least 10 times the rms thermal velocity \citep[see also][]{1992ApJ...399..551M}. 
Following \citet{2020ApJ...893..152R}, we estimate the one-dimensional turbulent velocity $\sigma_{\rm tur}$ for the tentative TSAS components by assuming that the nonthermal component of the velocity dispersion is entirely accounted for by turbulent motions such that 
\begin{equation}
\sigma_{\rm tur}^2 = \sigma_{v_0}^2-\sigma_{\rm th}^2=(\Delta v_0/2.355)^2 - k_bT_s/m_{\rm H_0},
\label{e:sig_tur}
\end{equation}
\noindent where $m_{\rm H_0}$ is the mass of hydrogen atom, $\sigma_{\rm th}$ is the one-dimensional rms thermal velocity, and the spin temperature $T_s$ is assumed to be a good approximation of the kinetic temperature for the CNM.
The turbulent velocity is far too small to confine the tentative TSAS, with $\sigma_{\rm tur}=0.2$ \kms and $\sigma_{\rm th}=0.8$ \kms (see also Table \ref{tab:comp_pro}).
If we instead adopt a minimum possible value of 10\,K for the spin temperature, we can calculate a lower limit for $\sigma_{\rm th}$ and the corresponding upper limit for $\sigma_{\rm tur}$ (as denoted by $\sigma^{T_s=10\,K}_{\rm th}=0.3$ \kms and $\sigma^{ T_s=10\,K}_{\rm tur}=0.5$ \kms\ in Table \ref{tab:comp_pro}).
Even in this case, the turbulent velocity 
is still less than 10 times the rms thermal velocity.
This suggests that the turbulent velocity is not sufficient to confine the tentative TSAS.

\section{Conclusions} \label{sec:conc}
We have obtained \hi\ absorption spectra in two epochs 0.36\,yr apart toward PSR~B1557$-$50 with the Parkes telescope.
We have detected one component with marginally significant absorption variations, which
probes astronomical-unit-scale atomic structure in the Milky Way.
Our main results are summarized as follows:

1. One TSAS component at $\sim$$-$54 \kms\ was marginally detected with an S/N of 3.2 after carefully reducing artifacts on the difference spectrum.

2. The characteristic plane-of-sky spatial scale of the TSAS component is 17\,au, with an inferred volume density (assuming a spherical cloud) of $2.7\times10^5$\,cm$^{-3}$, and thermal pressure of $2.1\times10^7 \rm\,cm^{-3}$\,K, suggesting that the TSAS is overdense and overpressured by two to three orders of magnitude.

3. The inferred overpressured TSAS has sufficiently high column density to be in thermal equilibrium with the ambient ISM, while still being overpressured.

4. The elongation factor
of the TSAS cloud would need to be $\sim$5000 in order for it to be in pressure equilibrium with a canonical CNM. This is much larger than expected from disks or cylinders \citep{1997ApJ...481..193H}, and than the aspect ratio of observed ISM filaments. 

5. In addition to optical depth variations, there is some evidence of line width and temperature variations in the CNM components observed at the two epochs.
This may hint at a 
possible formation mechanism
involving 
WNM collisions.

6. We consider a scenario in which the TSAS represents the tail-end of a turbulent cascade, and find that the turbulent line width would be insufficient to confine such an overpressured cloud.

Further monitoring of this pulsar, as well as other pulsars with previous \hi\ absorption measurements using single-dish telescopes (e.g. Parkes, FAST), would enable us to detect a larger population of TSAS, and with sensitivity to a wide range of spatial scales, which may be critical in better probing the role of turbulence in TSAS formation.
More generally, future high-resolution, interferometric observations (e.g. with the VLA, ALMA) of atomic and molecular lines (e.g. \hi, CO, C\textsc{i}, C\textsc{ii}, SiO) toward TSAS associated with a larger range of spin temperature estimates are necessary to better constrain TSAS cooling and heating mechanisms, and shed crucial light onto TSAS formation.

\acknowledgments
 This work is supported by National Natural Science Foundation of China (NSFC) program Nos.\ 11988101, 11725313, 11690024, by the CAS International Partnership Program No.\ 114-A11KYSB20160008, and the National Key R\&D Program of China (No.\ 2017YFA0402600). 
Cultivation Project for FAST Scientific Payoff and Research Achievement of CAMS-CAS.
 J.R.D. is the recipient of an Australian Research Council (ARC) DECRA Fellowship (project number DE170101086). 
 We are grateful to Lawrence Toomey for supporting the data transfer and processing. We thank CSIRO computing resources for data storage and processing.  We express our thanks to Claire Murray for Gaussian decomposition and fitting discussions; Shi Dai, Weiwei Zhu, Lei Zhang, and Chenchen Miao for dispersion measurements (DM) discussions; Jumei Yao for DM-based distance discussions; James Green for his help in observation arrangement at Parkes; Zhichen Pan and Yi Feng for supporting the data transfer; Lei Qian for the turbulent velocity discussion; Zheng Zheng for baseline removal discussion; Chaowei Tsai and Guodong Li for the MCMC discussion. 

\appendix

\section{The distance -- radial velocity curve Derivation} \label{sec:d_rv}
\citet{2019ApJ...870L..10M} used a sample of 773 Classical Cepheids with precise distances based on mid-infrared period–luminosity relations coupled with proper motions and radial velocities from Gaia to construct an accurate rotation curve of the Milky Way up to a distance of $\sim$20\,kpc from the Galactic center.
They found linear rotation curves describe that data much better than a simple constant rotation curve and are more consistent with previous observations than the universal rotation curve.

For an object in circular rotation about the galactic center at radius $R$ with circular velocity $\omega$, 
by adopting the best-fit linear model from \citet{2019ApJ...870L..10M}
\begin{equation}
\frac{\omega}{\omega_0}= 1.046\left( \frac{R_0}{R} \right)-0.046,
\label{e:blm}
\end{equation}
\noindent where $\omega_0$ is the angular velocity of the Sun's rotation around the Galaxy,
the radial velocity with respect to the LSR is given by 
\begin{equation}
V_{r}= \left[ 1.046\left(\frac{R_0}{R}\right) -0.046 \right] \times \Theta_0\sin l\cos b+V_{\Pi}\cos l\cos b,
\label{e:vr_R}
\end{equation}
\noindent where $R_0$ and $V_{\Pi}$ represent the galactocentric distance of the Sun and the net outward motion of the LSR with respect to the Galactic object that is 4.2 \kms,  respectively.
Similar to \citet{2008ApJ...674..286W}, we add and subtract velocities of 7 \kms to $V_{r}$ to estimate the uncertainties in distance limits due to streaming and random gas motions in the Galaxy.

With $R=(R_0^2+d^2-2R_0d\cos l)^{1/2}$, the distance--radial velocity relation can be derived
\begin{equation}
d= \pm \sqrt{\left(\frac{1.046R_0\Theta_0\sin l\cos b}{V_r-V_{\Pi}\cos l \cos b\pm7+1.046\Theta_0\sin l\cos b}\right)^2-R_0^2\left(1-\cos l^2\right)}+R_0\cos l.
\label{e:dis_vr}
\end{equation}

\section{Gaussian Decomposition and Spin Temperature Estimation} \label{sec:gs}

We largely followed the method described in \citet{2018ApJS..238...14M} to estimate the spin temperature. 
This method was developed based on the strategy first proposed by \citet{2003ApJS..145..329H}.
They used a least-squares fitting method to fit Gaussian components to both \hi\ absorption and emission spectra in order to estimate the physical properties of the \hi\ clouds.
Compared to the traditional fitting method described in \citet{2003ApJS..145..329H}, \citet{2018ApJS..238...14M} adopted an autonomous Gaussian decomposition algorithm (Gausspy), which implements a derivative spectroscopy technique to use supervised machine learning to estimate the number of Gaussian features and their properties.
We used an upgraded, fully autonomous, Gaussian decomposition algorithm via its open-source Python implementation \citep[GaussPy+;][]{2019A&A...628A..78R} to make initial guesses for Gaussian components toward PSR~B1557$-$50 and applied a least-squares fitting to refine the results.

According to  \citet{2003ApJS..145..329H}, the optical depth spectrum $\tau(v)$ along the LOS with a set of $N$ Gaussian components can be written as
\begin{equation}
\tau (v)= -\ln(I/I_0)= \sum_{n=0}^{N-1} \tau_{0,n} \cdot e^{- 4 \ln{2} \left(v-v_{0,n}\right)^2/ \Delta v_n^2},
\label{e:tau_abs}
\end{equation}

\noindent where $\tau_{0,n}$, $v_{0,n}$, $\Delta v_n$ are the the peak optical depth, central velocity and FWHM of the $n$th component.

The optical depth spectrum is only contributed by the CNM, while the brightness temperature-calibrated $T^{\rm off} (v)$ consists of both the CNM and the WNM,
\begin{equation}
T^{\rm off} (v) = T_{B, \rm CNM} (v) + T_{B, \rm WNM}(v),
\label{e:tb}
\end{equation}

\noindent where $T_{B, \rm CNM}$ is the $\rm \hi$ emission contributed by the CNM and $T_{B, \rm WNM}$ is the $\rm \hi$ emission contributed by the WNM. 
The H\textsc{i} emission contributed by $N$ CNM components can be written as

\begin{equation}
T_{B,\rm CNM}(v) = \sum_{n=0}^{N-1} T_{s,n} \left(1-e^{-\tau_n(v)}\right) e^{-\sum_{m=0}^M \tau_m(v)},
\label{e:em_CNM}
\end{equation}

\noindent where $m$ represents $M$ absorption clouds lying in front of the $n$th cloud, and $T_{s,n}$ represents the spin temperature for the $n$th component. For the H\textsc{i} emission contributed by the WNM $T_{B, \rm WNM}(v)$, can be considered as a set of $K$ Gaussian functions. For each $k$th component, there is a fraction $\mathscr{F}_k$ of the WNM is located in front of all CNM components. 

\begin{equation}
T_{B, \rm WNM}(v) = \sum_{k=0}^{K-1} \left[ \mathscr{F}_k + \left(1- \mathscr{F}_k\right)e^{-\tau(v)} \right] \cdot T_{0,k} e^{\frac{- 4 \ln{2} \left(v-v_{0,k}\right)^2}{ \Delta v_k^2}}.
\label{e:em_WNM}
\end{equation}
where $v_{0,k}$, $\Delta v_k$, and $T_{0,k}$ represent the central velocity, FWHM, and peak brightness temperature for the $k$th emission component.

Based on the method described in \citet{2018ApJS..238...14M},
spin temperatures can be estimated with the following steps:

\begin{enumerate}
\item Decompose Gaussian components for $\tau(v)$ with GaussPy+ to make initial guesses.
Following the method described in \citet{2019A&A...628A..78R}, we applied the algorithm on the optical depth profiles constructed from absorption lines of PSR~B1557$-$50. 
We chose the two-phase smoothing parameters $\alpha_1$ = 2.58 and $\alpha_2$ = 5.14, and a minimum S/N of 5 for signal peaks in the data to decompose the optical depth 
profiles.

\item Fit $N$ components to $\tau(v)$ via least-squares fitting using the {\tt\string LMFIT} package \citep{matt_newville_2019_3381550}. 
The components were selected from step (1), restricted to those with line widths less than 20 \kms, and which are well separated.
The mean velocities, widths, and amplitudes are allowed to vary by $\pm 20 \%$ with respect to the GaussPy+ fit parameters.
$\tau(v)$ was decomposed into 10 and 12 components for E1 and E2, respectively, which are shown as blue lines in the third panels of Figure ~\ref{f:final_all_fit}. 

\begin{figure*}
\begin{center}
\includegraphics[width=0.49\textwidth]{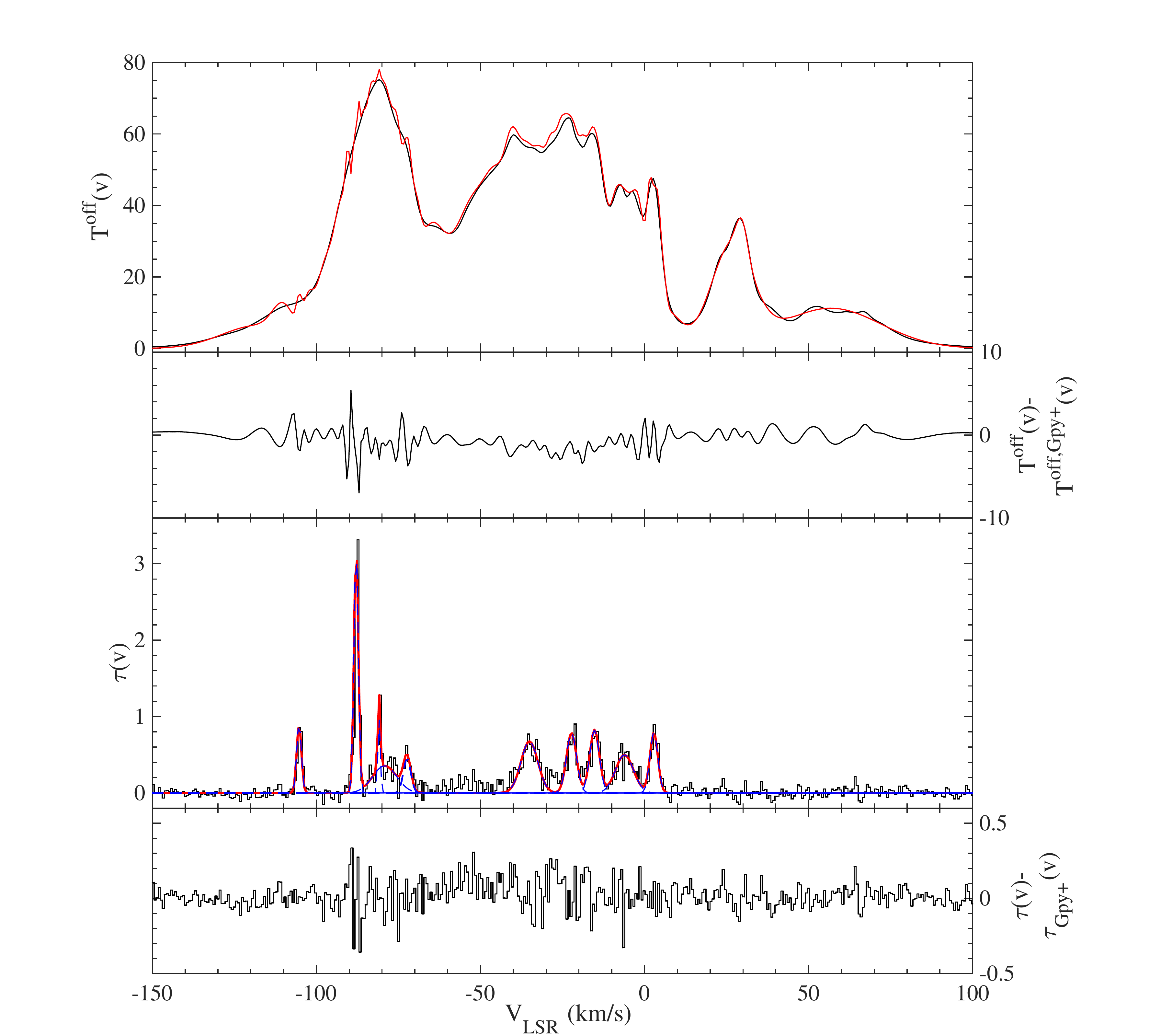}
\includegraphics[width=0.49\textwidth]{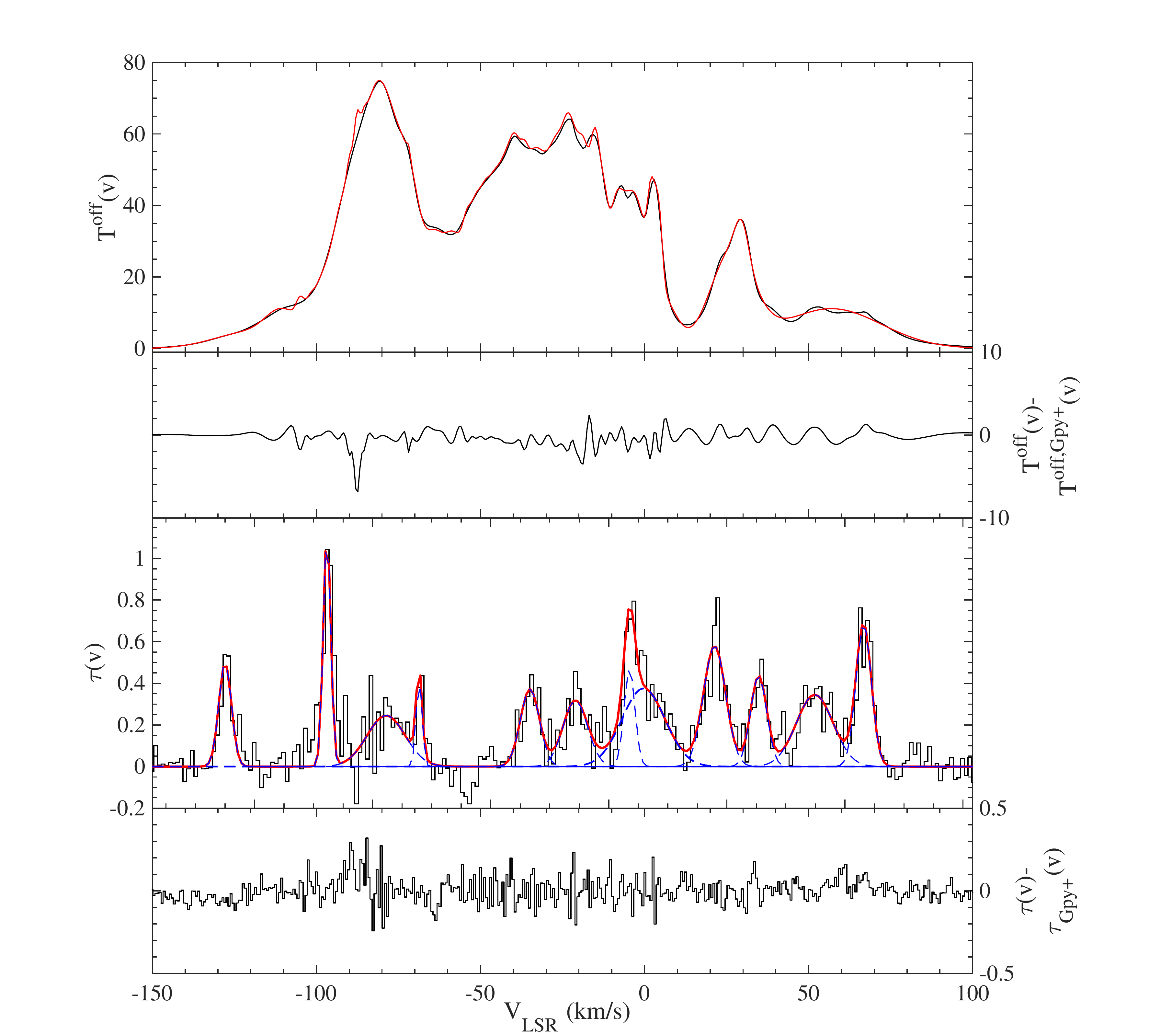}
\caption{For two epoch observations (left: E1, right: E2). The brightness temperature-calibrated \hi\ emission spectrum $T^{\rm off}(v)$ and the fitting spectra $T^{\rm off, Gpy+}(v)$ in the red line (the top panels), the fitting residual of the \hi\ emission spectrum (the top second panels), the optical depth profile $\tau(v)$, the decomposed Gaussian components in the dashed blue lines, and the fitting optical depth profile $\tau_{Gpy+}(v)$ in the red line (the top third panels), and the fitting residual of the optical depth profile (the bottom panels). The spectra were binned in velocity to a channel width of 0.6 \kms\ . }
\label{f:final_all_fit}
\end{center}
\end{figure*}

\item Fit the $N$ components from $\tau(v)$ to $T^{\rm off}(v)$ via least-squares fitting. 
The mean velocities and widths are allowed to vary by $\pm10\%$. 
$T_s$ are constrained to between $0<  T_{s,n} \leq T_{k,{\rm max}, n} = 21.866\cdot \Delta v_n^2$, to produce physically realistic spin temperatures.

\item Subtract the best-fit model in step (3) from $T^{\rm off}(v)$ to produce a residual emission spectrum, which contains only
WNM components not previously modeled.

\item Fit $K$ components to the residual emission spectrum from (4) with GaussPy+, using $\alpha_1$ = 2.58, $\alpha_2$ = 5.14, and S/N=740. 

\item Use least-squares fitting to fit $T^{\rm off}(v)$ with $N+K$ Gaussian components from steps (2) and (5). 
The mean velocities and widths are allowed to vary by $10\%$ with respect to the previously fitted values. The amplitudes are constrained so that $T^{\rm off}(v) > 0$. 
The final estimation of $T_s$ for the $N$ absorption components and the Gaussian parameters of the $K$ emission-only components is computed based on Equations on \ref{e:tb}--\ref{e:em_WNM}. 
\end{enumerate}

When the absorption components overlap in velocity significantly, the order of each component will affect $T_{B,\rm CNM}(v)$ (\citealt{2003ApJS..145..329H,2010ApJ...720..415S,2018ApJS..238...14M}). 
There are a maximum of $N!$ different orderings of components along the LOS.
In the case of our optical depth profiles, the components are well separated, so 
we did not need to consider the ordering phenomenon during fitting.
Previous studies have indicated that the values of $\mathscr{F}_{k}$ have significant effect on the derived spin temperature (\citealt{2003ApJS..145..329H,2010ApJ...720..415S,2018ApJS..238...14M}).
We followed previous analyses to estimate the spin temperature by assigning value of 0.0, 0.5, or 1.0 to $\mathscr{F}_k$.
Therefore, there are three possible cases for the final fit of $T^{\rm off}(v)$ from our data.

The final spin temperatures are calculated by estimating the weighted mean and standard deviation over the three iterations following \citealt{2003ApJS..145..329H}.
For the spectra observed in E1 and E2, we fit 10 ($N$ =10) and 12 ($N$ =12) absorption components and 27 ($K$ =27) and 26 ($K$ =26) emission-only components, respectively.
Figure \ref{f:final_all_fit} illustrates the Gaussian decomposition and fitting of the \hi\ emission and optical depth profiles for E1 and E2, respectively.
The main results of the CNM decomposition are shown in Table \ref{tab:params}.

\bibliography{sample63}{}  
\bibliographystyle{aasjournal}



\end{document}